\documentclass[preprint,preprintnumbers,amsmath,amssymb,color]{revtex4}

\usepackage{graphicx}
\usepackage{bm}

\def\half{{1\over 2}}

\def\o1{{\mathrm{o}(1)}}
\begin{document}

\preprint{CAS-KITPC/ITP-210}

\preprint{USTC-ICTS-10-19}

\title{~\\ \vspace{2cm}
A New Entropic Force Scenario and Holographic Thermodynamics \vspace{1cm}}
\author{Wei Gu}\email{guwei@mail.ustc.edu.cn}
\author{Miao Li}\email{mli@itp.ac.cn}
\author{Rong-Xin Miao}\email{mrx11@mail.ustc.edu.cn}
\affiliation{Kavli Institute for Theoretical Physics, Key Laboratory
of Frontiers in Theoretical Physics, Institute of Theoretical
Physics, Chinese Academy of Sciences, Beijing 100190, People's
Republic of China, Interdisciplinary Center for Theoretical Study,
University of Science and Technology of China, Hefei, Anhui 230026,
China\vspace{2cm}
}%

\begin{abstract}
We propose a new holographic program of gravity in which we introduce a surface
stress tensor. Our proposal differs from Verlinde's in several aspects. First,
we use an open or a closed screen, a temperature is not necessary but a surface
energy density and pressure are introduced. The surface stress tensor is proportional
to the extrinsic curvature. The energy we use is Brown-York energy and the
equipartiton theorem is violated by a non-vanishing surface pressure.
We discuss holographic thermodynamics of a gas of weak gravity and find a chemical potential,
and show that Verlinde's program does not lead to a reasonable thermodynamics.
The holographic entropy is similar to the Bekenstein entropy bound.

\end{abstract}


\maketitle

\section{Introduction}

That Gravity may not be a fundamental force has long been  suspected, and gained more credit and
interest since Verlinde's work \cite{Verlinde} (see also \cite{Pad}). Verlinde proposed that gravity
is not only an emergent force, but specifically an entropic force. His work builds upon an early
work of Jacobson \cite{Jacobson}, and there have appeared many following up papers afterwards \cite{enfor}.

The fact that gravity is emergent is not surprising, given the tremendous progress in AdS/CFT
in the last decade. This correspondence tells that at least for an anti-de Sitter space or a related
space-time, quantum gravity can be fully described by a quantum field theory without gravity
in a lower dimension. It is certainly true that gravity remains a fundamental force in the bulk
description, but this theory may be less fundamental than the corresponding holographic field
theory, since the latter has a well-developed understanding including its definition and quantization.
Some would like to say that the holographic field theory offers a nonperturbative defintion
of quantum gravity in the bulk.

The program of Verlinde is more general than gauge/gravity correspondence, however it is also rather primitive,
since we have  little information about the theory on a holographic screen. The main
ingredients in Verlinde's scenario including a temperature, the equipartition theorem of energy,
and an entropy increase when a test particle approaches the holographic screen. However,
the temperature depends on the local bulk geometry of the full space-time, and we do not know
how this relation arises. Moreover, the equipartition sounds very peculiar.

In this paper, we shall challenge Verlinde's definition of energy, and the assumption of
the equipartition theorem. The main reason is that Verlinde's energy does not give us a well-behaved
thermodynamics, specifically, the holographic entropy of a bulk gas computed using Verlinde's
definition of energy has an density independent area term. We propose to use Brown-York semi-local energy
to replace Verlinde energy. In doing so, we introduce a surface stress tensor on a holographic screen,
this tensor give us both the surface energy density as well as the surface pressure-this is missing
in Verlinde's program.

The surface stress tensor is proportional to the extrinsic curvature of the screen in space-time.
This may hint that we need more geometric data than Verlinde does. However, in deriving the
Einstein equations, only two physical quantities are involved, namely the surface energy density
and the surface pressure, compared with Verlinde temperature, we need one more scalar. This may
not be a disadvantage, rather it is an advantage since we know more about what is living on the
screen. We do not have to assume the equipartition, and we show that the equipartition rule
is violated by a non-vanishing surface pressure.

Another difference between our proposal and Verlinde's is that we use an open or closed screen, while
Verlinde uses a closed one. We do not need a temperature, though in discussing holographic
thermodynamics we adopt Verlinde temperature. The use of a closed screen in Verlinde's program
introduces the problem of possible negative temperature, thus his derivation of the Eintein
equations is incomplete. In his derivation, a total energy on the closed screen is needed, while in
our derivation we need to study the energy flow through an open/closed surface. Thus our derivation
is similar to Jacobson's original derivation \cite{Jacobson}, the difference is that our holographic
screen is time-like and his is null, so we have more data on the screen since the surface pressure
always vanishes on a null screen.

One of the present authors and Yi Pang proved a no-go theorem some time ago \cite{lipang}, stating
that it is impossible to accommodate a high derivative theory of gravity in Verlinde's program.
It appears that there is no such a no-go theorem in our program, we may replace the extrinsic
curvature as the surface stress tensor by one involving higher derivatives too to derive equations
of motion in a high derivative theory.

We also study holographic thermodynamics of a spherically symmetric and static system using our
program. One derives one more important quantity, the chemical potential, for this system.
We obtain a very interesting result: The holographic entropy of a gas of weak gravity has a
form parametrically similar to the Bekenstein entropy bound. We take this as an indication that
our program is correct. We also show that there is always a density independent area term
in the holographic entropy of the gas if we use Verlinde's program. Note that, the holographic
entropy is usually much larger than the statistical entropy of the gas, we explain this
fact by the tremendous contribution to entropy from gravity. It has been known for a long time
that the usual statistical entropy of a collapsing system violates the second law of
thermodynamics, our explanation is that we need to use the holographic entropy rather
than the usual statistical entropy. We expect that the holographic entropy smoothly
crosses-over to the black hole entropy in a black hole formation process.

It remains to derive a general, geometric formula for the chemical potential for a general
system, we leave this to a future work.

The plan of this paper is the following. We explain our holographic derivation of the
Einstein equations in sect.2, and compare our proposal to those of Verlinde and Jacobson
in sect.3. We discuss holographic thermodynamics of a gas of weak gravity in sect.4 and
show that Verlinde's program does not give us a reasonable answer to this problem in
sect.5. We concldue in sect.6. The appendix gives the solution of a gas of weak
gravity to the second order in $G$.

\section{The Einstein Equations and holographic screens}

In this section, we will derive the Einstein equations from the
holographic principle in a different way from Verlinde's.

Let us first briefly review Verlinde's derivation of the Einstein
equations. There are a few important ingredients in Verlinde's scenario. The first is
a time-like holographic screen and a temperature on it. This temperature is motivated
by Unruh temperature, and plays an imperative role in the derivation of the Newton's law
of gravitation. We will follow Verlinde to consider a time-like holographic screen and
his definition of temperature on it, however, we shall not assume that the screen is closed,
not to mention an equipotential one. A time-like holographic screen in our proposal can
be either closed or open, and does not have to be an equipotenttial surface so long if
the temperature on the surface is positive everywhere. In fact, in the following derivation
of the Einstein equations, we do not even need a temperature, the temperature is
introduced only when we discuss holographic thermodynamics in Sect.4.  For Verlinde, a closed surface can not be any closed
surface since most of time the temperature may become negative.

The second ingredient of Verlinde's proposal is the equipartition theorem, namely, assuming the
number of degrees of freedom $N$ be proportional to the area, the energy of this area is $TN/2$. This is a peculiar
assumption, since in statistical mechanics it is valid only for a gas with high temperature without
interaction. We will not assume this. The third ingredient is that the holographic energy on the
screen is equal to the Tolman-Komar mass, we will also differ from Verlinde for this point. The reason for us to abandon
the Tolman-Komar mass is that this definition does not offer us a consistent thermodynamics on
the holographic screen.

To define Verlinde temperature, we need the definition of a generalization of Newton's potential
\cite{wald}
\begin{equation}\label{np}
\phi=\half\log(-\xi^a\xi_a),
\end{equation}
where $\xi^a$ is a local time-like Killing vector. The definition of Verlinde temperature is
\begin{equation}\label{vt}
T={\hbar\over 2\pi}e^\phi N^a \nabla_a \phi,
\end{equation}
where $N^a$ is a vector normal to the time-like surface $\Sigma$ (this is a 2+1 dimensional surface
whose constant time slice is $B$), whose precise definition will be given shortly. Next, the number of
degrees of freedom on an area $dA$ of $B$ is $dN=dA/(G\hbar )$, so according to the equipartition assumption
the energy on $B$ is
\begin{equation}\label{eb}
E=\half \int_B TdN,
\end{equation}
after some manipulations, this is equal to $1/(4\pi G)\int_V R_{ab}n^a\xi^b dV$, where $V$ is the volume enclosed by $B$ and
$n^a$ is normal to $V$(different from $N^a$).
If one takes this to be equal to the Tolman-Komar mass $2\int_V (T_{ab}-\half Tg_{ab})n^a\xi^b dV$,
one may see that the Einstein equations appear in an integral form.

Of course the above is not a complete derivation of the Einstein equations. Let us leave aside the issue
whether $\xi^a$ and $n^a$ can be arbitrary and independent of each other.
To derive the differential equations from the integral equation, $V$ must be arbitrary and arbitrarily
small, this can not be the case since to have the temperature $T$ always positive on $B$, the normal
vector $N^a$ can not reverse its direction.

We now turn to our derivation of the Einstein equations. Our
proposal is mainly motivated by the work of Brown and York
\cite{Brown} and that of Jacobson \cite{Jacobson}. We shall use the
semi-local energy of Brown and York rather than the Tolman-Komar
mass, this is a good definition with a nice thermodynamics on
holographic screens. We shall use an open screen (we will not
introduce a temperature in this section, so it can be closed), this
is motivated by Jacobson's work which was the source of Verlinde's
work after all. In \cite{Jacobson}, Jacobson used a null open
screen, here we will use a time-like open screen. A time-like
holographic screen encodes much more information than a null one, as
we shall see later.

To begin, let us start with some definitions. Our holographic screen is a 2+1 dimensional time-like hypersurface $\Sigma$,
which can be open or closed, and we will work on a patch of it, $M$ is the 3+1 dimensional
space-time.  We use $x^a$, $g_{ab}$, $\nabla_{a}$, $y_i$,
$\gamma_{ij}$ and $D_{i}$ (here $a$,$b$ run from 0 to 3, and $i$,$j$
run from 0 to 2) to denote the coordinates, metric and covariant
derivatives on $M$ and $\Sigma$, respectively.
\begin{figure}
\includegraphics[scale=0.36]{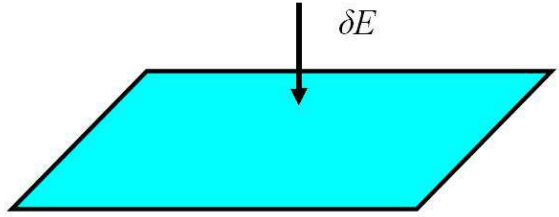}
\caption{An energy flux $\delta E$ passing through an open patch on
the holographic screen. } \label{fig1}
\end{figure}

Similar to Jacobson's idea \cite{Jacobson}, we consider an energy
flux $\delta E$ passing through an open patch on the holographic
screen $d\Sigma=dAdt$ (see FIG. \ref{fig1})
\begin{equation}\label{E1}
 \delta E=\int_{\Sigma} T_{ab}\xi^{a}N^{b}dA dt,
\end{equation}
where $T_{ab}$ is the
stress tensor of matter in the bulk $M$, $\xi^a$ is the Killing
vector, and $N^a$ is the unit vector normal to $\Sigma$. To define
the normal $N^a$ we may assume $\Sigma$ be specified by a function
$f_{\Sigma}(x^{a})$ with $x^a$ subject to
\begin{equation}\label{fSigma}
f_{\Sigma}(x^{a})=c .
\end{equation}
Obviously, a vector normal to $\Sigma$ is
proportional to $g^{ab}\nabla_{b}f$. Thus, after normalization it
follows that
\begin{equation}\label{Na}
N^{a}=\frac{g^{ab}\nabla_{b}f_{\Sigma}}{\sqrt{\nabla_{b}f_{\Sigma}\nabla^{b}f_{\Sigma}}} .
\end{equation}

According to the holographic principle, every physical
process happening  in the bulk corresponds to a process on the
holographic screen. The above energy flux Eq.(\ref{E1}) represents change of energy
on one side of the holographic screen, it is then natural to assume
that it is equal to the change of energy on the patch of screen through which
the bulk energy flows.

To calculate the change of energy on the screen, we need to introduce the surface
energy density $\sigma$ on the screen or more generally the surface stress tensor $\tau_{ij}$.
The relations between the surface energy density, the surface energy flux $j$  and the surface
stress tensor are
\begin{equation}\label{sigma}
 \sigma=u_{i}\tau^{ij}\xi_{j}\ ,\ j=-m_{i}\tau^{ij}\xi_{j},
\end{equation}
where $u_{i}$, $m_{i}$ are the unit vectors normal to the screen's
boundary $\partial\Sigma$ (we will give the expressions of $u_i$ and
$m_i$ under Eq.(\ref{Mi})), $\xi_{i}$ is the Killing vector on
$\Sigma$. In quasi-static spacetime $u^{i}$ is related to $\xi^{i}$
by $u^{i}=e^{-\phi}\xi^{i}$, and $\phi$ is the Newton's potential
defined by $\phi=\half\log(-\xi^i\xi_i)$ on the screen. Note that
$\sigma$ and $j$ are the energy density and energy flux on the
screen measured by the observer at infinity. Apparently, central to
our discussion is the choice of $\tau_{ij}$, we will ultimately use
the Brown-York expression, but for now let us be more general.
Naturally, $\tau^{ij}$ should depend only on geometry of the
boundary, thus we assume the following
\begin{equation}\label{t}
 \tau^{ij}=n (K^{ij}- K\gamma^{ij}),
\end{equation}
where $n$ is a constant to be determined, and $K^{ij}$ is the
extrinsic curvature on $\Sigma$ defined by
\begin{equation}
 K_{ij}=-e^{a}_{i}e^{b}_{j}\nabla_{a}N_{b},
\end{equation}
where $e^{b}_{j}=\frac{\partial x^{a}}{\partial
 y^{i}}$ is the projection operator satisfying $N_{a}e^{a}_{i}=0$.
It should be stressed that Eq.(\ref{t}) is the key ansatz in our
scheme to derive the Einstein equations. We also note that one may
replace the coefficient of the term $K$ in Eq.(\ref{t}) by any other
number, this does not affect the derivation of the Einstein equations.

On the screen, the change of energy has two sources, one due to
variation of the energy density $\sigma$, another is due to energy
flowing out the patch to other parts of the screen, given by the
energy flux $j$. The energy variation on the patch is then given by
\begin{equation}\label{EE2}
\delta E=\int(u_{i}\tau^{ij}\xi_{j})dA|^{t+dt}_{t}-\int
m_{i}\tau^{ij}\xi_{j}dydt=-\int_{\partial\Sigma}(M_{i})\tau^{ij}\xi_{j}\sqrt{h}dz^{2}=-\int_{\Sigma}D_{i}\tau^{ij}\xi_{j}dAdt ,
\end{equation}
\begin{figure}
\includegraphics[scale=0.36]{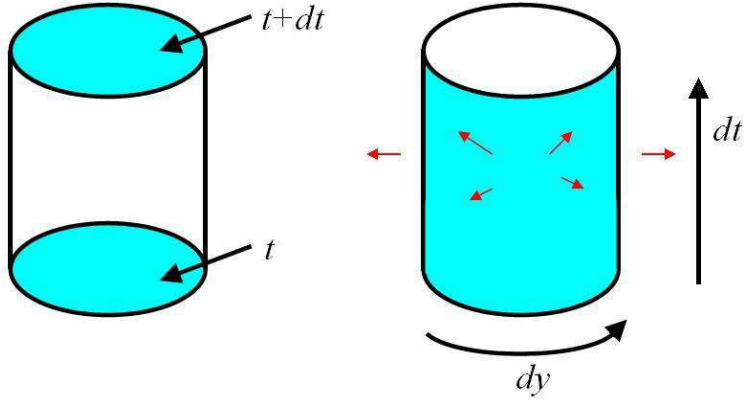}
\caption{Left panel: The first term
$\int(u_{i}\tau^{ij}\xi_{j})dA|^{t+dt}_{t}$ in Eq.(\ref{EE2}) is due
to change of the density. Right panel: The second term $-\int
m_{i}\tau^{ij}\xi_{j}dydt$ is the energy flow (denoted by the red
arrows) through the patch boundary parameterized by $y$. }
\label{fig2}
\end{figure}
where the first term in the first equality is due to change of the
density and the second term is the energy flow through the patch
boundary parameterized by $y$ (see Fig. \ref{fig2}). These two terms
can be naturally written in a uniform form, since the boundary of
the patch consists of two space-like surfaces ($d\Sigma$ at $t$ and
$t+dt$), and a time-like boundary. $h$ is the determinant of the
reduced metric on $\partial\Sigma$, $D_{i}$ is the covariant
derivative on $\Sigma$. $M_{i}$ is a unit vector in $\Sigma$ and is
normal to $\partial\Sigma$. Let us choose a suitable function
$f_{\partial\Sigma}(y^{i})=c$ on
 $\Sigma$ to denote the boundary $\partial\Sigma$, then we have
\begin{equation}\label{Mi}
M^{i}=\frac{\gamma^{ij}D_{j}f_{\partial\Sigma}}{\sqrt{D_{j}f_{\partial\Sigma}D^{j}f_{\partial\Sigma}}}.
\end{equation}
Notice that when $M^{i}$ is along the direction of $dy^{0}$($dt$) it
becomes $u^{i}$, and when along the direction of
$dy^{i}$($dy^{1},dy^{2}$) it becomes $m^{i}$.

From Eq.(\ref{t}) and the Gauss-Codazzi equation $(R_{ab}-R
g_{ab}/2)N^{a}e^{b}_{i}=-D_{j}(K^{j}\ _i-K\gamma^{j}\ _i)$, one can
rewrite Eq.(\ref{EE2}) as
\begin{equation}\label{E2}
\delta
E=\int_{\Sigma}n(R_{ab}-\frac{R}{2}g_{ab})\xi^{i}e^{a}_{i}N^{b}dAdt.
\end{equation}
According to the holographic
principle if we equate Eq.(\ref{E1}), the energy flow through the patch of the holographic screen
$\Sigma$, and Eq.(\ref{E2}), the energy change on this patch, we will obtain the Einstein equations.

In the following derivations, for simplicity we will focus only on
a quasi-static process. Recalling that in Eq.(\ref{fSigma}) we use
$f_{\Sigma}(x^{a})=c$ to denote the holographic screen $\Sigma$. In
the quasi-static limit, $f_{\Sigma}(x^{a})$ is independent of time,
so
$N_{a}\sim(0,\partial_{1}f_{\Sigma},\partial_{2}f_{\Sigma},\partial_{3}f_{\Sigma})$.
Note that $\xi^{a}\simeq(1,0,0,0)$ in the quasi-static limit, thus
we have $N_{a}\xi^{a}\rightarrow 0$. The Killing vector
$\xi_{i}$ on $\Sigma$ is induced from the Killing vector $\xi_a$
in the bulk $M$:
\begin{equation}
 \xi_{i}=\xi_{a}e^{a}_{i},\ \
 D_{(i}\xi_{j)}=\nabla_{a}(\xi_{b}-N_{c}\xi^{c}N_{b})e^{a}_{(i}e^{b}_{j)}=K_{ij}N_{a}\xi^{a}\rightarrow
 0 .
\end{equation}
Thus, in the quasi-static limit, we can rewrite Eq.(\ref{E1}) as
\begin{equation}\label{E11}
 \delta E=\int_{\Sigma} T_{ab}\xi^{i}e^{a}_{i}N^{b}dA dt.
\end{equation}
Equating Eq.(\ref{E11}) and Eq.(\ref{E2}) results in
\begin{equation}\label{Einstein}
n(R_{ab}-\frac{R}{2}g_{ab})e^{a}_{i}N^{b}=T_{ab}e^{a}_{i}N^{b}.
\end{equation}
Note that $g_{ab}e^{a}_{i}N^{b}=0$, thus we can add a cosmological
constant term $\Lambda g_{ab}$ to the left of the above equation.
Defining the Newton's constant as $G=\frac{1}{8\pi n}$, we then get
the Einstein equations
\begin{equation}\label{Einstein1}
R_{ab}-\frac{R}{2}g_{ab}+\Lambda g_{ab}=8\pi G T_{ab}.
\end{equation}
Substituting $n=\frac{1}{8\pi G}$ into Eq.(\ref{t}), we obtain
\begin{equation}
\tau^{ij}=\frac{1}{8\pi G}(K^{ij}- K\gamma^{ij}),
\end{equation}
this is just the quasi-local stress tensor of Brown-York defined in \cite{Brown}.
As emphasized in \cite{Brown}, $\tau^{ij}$ characterizes the entire
system, including contributions from both the gravitational field
and the matter fields. In addition, generally $\tau^{ij}$ need to be
subtracted in order to derive the correct energy. We will discuss
this issue in the next section.

Note that, if we replace the extrinsic curvature in $\tau_{ij}$ in Eq.(\ref{t}) )by the combination
of the extrinsic curvature and higher derivative terms, we may derive the equations of
motion in a high derivative theory, thus avoiding the no-go theorem of \cite{lipang}.

To find the relationships between the bulk energy and the energy on
the holographic screen, let us consider the case of a small closed screen.
Using Eq.(\ref{E1}) and Eq.(\ref{EE2}), together with the
conversation of energy in the bulk
\begin{equation}\label{energy}
\int\sqrt{-g}\nabla_{a}(T^{ab}\xi_{b})dx^{4}=-(U_{a}T^{ab}\xi_{b})dV\mid^{t+dt}_{t}+\int_{\Sigma}
T_{ab}\xi^{a}N^{b}dA dt=0,
\end{equation}
we obtain
\begin{equation}\label{bb}
(u_{i}\tau^{ij}\xi_{j})dA|^{t+dt}_{t}=(U_{a}T^{ab}\xi_{b})dV\mid^{t+dt}_{t},
\end{equation}
where we have used the fact that a closed screen has no time-like boundary, thus there
is no the second term in Eq.(\ref{EE2}). The above relation describes the relationship between the bulk energy and the
energy on the holographic screen. Clearly, they are different from
each other at most by a time independent constant. This constant is
related to the subtraction scheme of the energy on the screen, and
generally it depends on the position of the screen. Since the
constant is independent of time, the subtraction scheme does not
affect our derivation of the Einstein equations. One may view Eq.(\ref{bb}) as a consequence
of the Einstein equations.

In all derivations of the Einstein equations through the holographic principle, two
data sets are involved, one set is the bulk data, another set is the screen data.
In our derivation, the important screen data are the surface stress tensor, while in Jacobson's
derivation, they are the entropy and Unruh temperature. In Verlinde's version, they
are Verlinde temperature and the equipartition assumption. In the next section, we compare our
proposal to Jacobson's as well as Verlinde's.

\section{Comparison with Jacobson and Verlinde}

The energy on a holographic screen is completely determined by the
surface energy density $\sigma$, as in Eq.(\ref{sigma}) and
Eq.(\ref{t}). We will show that this is different from Verlinde's
simple postulate $\half TA/G\hbar$, thus the quasi-local energy of
Brown-York is different from that of Verlinde. Nevertheless, a part
of the quasi-local energy surface energy is similar to that of
Verlinde.

 If we ignore the subtraction of the energy on the screen, we find
 in the static space
\begin{equation}\label{qle}
 E=\int(u_{i}\tau^{ij}\xi_{j})dA=\int
 \frac{1}{8\pi G}u_{i}(K^{ij}-K\gamma^{ij})\xi_{j}dA=\int \frac{T}{4G \hbar}dA+\int\frac{1}{8\pi G}K\exp(\phi)dA .
\end{equation}
In the derivation of the above equation the following formulas have
been used
\begin{equation}
\frac{2\pi
T}{\hbar}=N^{b}U^{a}\nabla_{a}\xi_{b}=N^{b}u^{i}e^{a}_{i}\nabla_{a}\xi_{b}=N_{a}u^{i}(e^{a}_{j}D^{j}\xi_{i}+N^{a}K_{ij}\xi^{j})=u^{i}K_{ij}\xi^{j} ,
\end{equation}
\begin{equation}
u_{i}\gamma^{ij}\xi_{j}=-\exp[\phi],
\end{equation}
where $\phi$ is the Newton's potential, and $U^a=e^{-\phi}\xi^a$ \cite{Verlinde}. We see that the first term in
Eq.(\ref{qle}) is precisely half of Verlinde's energy, but the second term is not.

Let the surface stress tensor of Brown-York assume the perfect fluid form
\begin{equation}\label{pf}
\tau_{ij}=e^{-\phi}(\sigma+p)u_iu_j+e^{-\phi}p\gamma_{ij}.
\end{equation}
$\sigma$ and $p$ are the the energy density and pressure on the
screen measured by the observer at infinity which are related with
the local energy density and pressure on the screen by a red-shift
$e^{-\phi}$. One can check that $T\sim \sigma +2p$ and $K\sim \sigma
-2p$ , thus our quasi-local energy agrees with Verlinde's energy
only when the surface pressure $p=0$, this is not generally valid,
and we shall see that $p$ will play an important role in holographic
thermodynamics. More precisely $TA/(2G \hbar)=(\sigma +2p)A$, we
rewrite the energy on the screen as
\begin{equation}
 E=\int \frac{T}{2G\hbar}dA-2\int p\ dA.
\end{equation}
If we are to insist $T$ be the physical temperature on the holographic screen, then the above
formula tells us that the first term is the result of the equipartition theorem, if the second
term vanishes. A non-vanishing pressure represents a deviation from the equipartition theorem.
Thus, only a system of surface dust satisfies the equipartition theorem, or more practically,
a system of motionless spin degrees of freedom.

To repeat what we briefly mentioned in the previous section, Verlinde derives the Einstein equations from
assumptions of the equipartition rule and the Tollman-Collman Mass
\begin{equation}\label{rvd}
 M=\frac{TN}{2}=\frac{TA}{2G\hbar}=\frac{1}{4\pi
 G}\int_V R_{ab}n^{a}\xi^{b}dV,
\end{equation}
\begin{equation}
 M=2\int_V (T_{ab}-\frac{T}{2}g_{ab})n^{a}\xi^{b}dV ,
\end{equation}
where $V$ is a spacelike hypersurface denoting the space volume,
$n^{a}$ is the unit vector normal to $V$. What should be mentioned
is that,  the Stokes theorem must be  applied in order to get a
volume integral from a surface integral $\frac{1}{2G\hbar}\int TdA$
in the first line of Eq.(\ref{rvd}). This requires that the screen
should be closed. However, as we said already, a negative
temperatiure problem will arise since for an arbitrary closed screen
Verlinde temperature can not keep positive in general. Imagine an
arbitrary small surface $B$, if $T$ is positive, then it becomes
negative when one goes around to the opposite side, the reason is
that the sign of $N^{a}$ will change on the opposite side of a
closed screen, leading to the change of the sign of the temperature,
which is related to $N^a$ by $T=\frac{\hbar}{2\pi
}N^{a}\nabla_{a}e^{\phi}$.

While in our proposal, we do not need a temperature. When we wish to keep the virtue of Verlinde's
derivation of the entropic force, we can consider an open screen on which $T$ is always positive. We also
abandon the unnatural assumption of
equipartition rule for a thermodynamics system on the holographic screen we know little about.

Finally, we compare our proposal with that of Jacobson \cite{Jacobson}. We were actually motivated by
mimicking his discussion for an open time-like screen, rather than a null one. In Jacobson's discussion,
he starts with an equation similar to Eq.(\ref{E1})
\begin{equation}
 \delta Q=\int_{H} T_{ab}\xi^{a}k^{b}dA d\lambda ,
\end{equation}
(H denotes the null surface, $\lambda$ is the affine parameter and
$k^a=\frac{dx^a}{d\lambda}$ is the tangent vector to $H$.)
 and then assumes that the energy flow is given by
the variation of energy on the null surface which in turn is given
by $T\delta S$ according to the first law of thermodynamics.
Assuming $dS=1/4 \delta A$, he arrives at the Einstein equations with
a undetermined cosmological constant. The role of $T\delta S$ is
played by the first term in Eq.(\ref{EE2}). On null surface we
replace $N^a$ and $u^i$ by $k^a$ and $l^a$ respectively, where
$l^{a}$ is a vector on $H$ satisfying $l^ak_a=-1, l^al_a=0$. Assume
$\tau_{ab}=nK_{ab}=n\nabla_ak_b$, we can derive
\begin{eqnarray}
 \delta
 Q&=&\int(l_{a}\tau^{ab}\xi_{b}dA)|^{d\lambda}_{0}=n\int(l^{a}K_{ab}\xi^{b}dA)|^{d\lambda}_{0}\nonumber\\
 &=&n\int (l^{a}\nabla_{a}k_b\xi^{b}dA)|^{d\lambda}_{0}=n\int (-l^{a}k^{b}\nabla_{a}\xi_bdA)|^{d\lambda}_{0}\nonumber\\
 &=&\frac{2\pi n}{\hbar}\int (T dA)|^{d\lambda}_{0}=\frac{(8\pi
 n)}{\hbar}T\delta(\frac{dA}{4})=T\delta S.
\end{eqnarray}
In the above derivations, we have used the formulas
$T=\frac{\hbar}{2\pi}l^{a}k^{b}\nabla_{b}\xi_a$, $k_{a}\xi^a=0$ and
$n=\frac{1}{8\pi G}$. On the null surface, the second term in
Eq.(\ref{EE2}) always vanishes, since there is no energy flow along
the null surface, this also means $p=0$ on the null surface. $p=0$
explains why the first law can be written in a simple form $\delta
E=T\delta S$.

To summarize, our proposal differs from Verlinde's proposal in that
it works also for an open screen, thus it is more appropriate for
derivation of the Einstein equations since a patch of open screen
can be arbitrarily small. We do not have to assume the equipartition
theorem of energy. More importantly, as we shall see, the knowledge
about $p$ is useful for discussing holographic thermodynamics (See
Table.\ref{table1} for details). Finally, Jacobson's derivation is
similar to ours but there is also no information about p (which
vanishes on a null screen), thus his program contains little
information about details of thermodynamics (see Table.\ref{table2}
for details).
\begin{table}
\caption{A comparison between Verlinde's and our proposals}
\begin{center}
\label{table1}
\begin{tabular}{|c|c|}
  \hline\hline
  ~~~Verlinde's proposal~~~   & ~~~~~~~~Our proposal~~~~~~~  \\
  \hline\hline
  ~~~Closed holographic screen~~~ & ~~~~~~~~Open or closed time-like screen~~~~~~~  \\
  \hline
  ~~~Temperature T~~~ & ~~~~~~~~ Without or with T~~~~~~~  \\
  \hline
  ~~~Tolman-Komar mass  ~~~ & ~~~~~~~~Brown-York energy~~~~~~~  \\
  \hline
  ~~~Equipartition  ~~~ & ~~~~~~~~Surface stress tensor~~~~~~~  \\
  \hline\hline
\end{tabular}
\end{center}
\end{table}

\begin{table}
\caption{A comparison with Jacobson's and our proposals}
\begin{center}
\label{table2}
\begin{tabular}{|c|c|}
  \hline\hline
  ~~ Jacobson's proposal ~~~~~& ~~~~~~~~Our proposal~~~~~~~  \\
  \hline\hline
  ~~~~~~~  Open null screen~~~~~~~~ & ~~~~~~~~Open or closed time-like screen~~~~~~~  \\
  \hline
  ~~~~T only  ~~~~~~~~ & ~~~~~~~~    T, p, $\mu$  ~~~~~~~  \\

  \hline
  ~~~~First law: $dE=TdS$  ~~~~~~~~ & ~~~~~~~~First law: $dE=TdS-pdA+\mu dN_f$~~~~~~~  \\
  \hline\hline
\end{tabular}
\end{center}
\end{table}

\section{Holographic Thermodynamics and the entropy bound}

We have seen in the previous two sections that not only we can derive the Einstein equations
of gravity, we also have an opportunity to discuss holographic thermodynamics, since we
have almost the ingredients, energy, temperature, pressure. We shall see, the final ingredient
will be the chemical potential $\mu$.

Let us for simplicity consider the situation in which Verlinde temperature on the holographic
screen is the same everywhere. Since the number of degrees of freedom plays no role in our proposal,
we reserve $N_f$ for the number of particles on the screen, (whatever it is, there may be
no particles but spin degrees of freedom etc.) if there is such a thing ($N$ will denote
the lapse function in the ADM formalism). The first law of thermodynamics on the holographic screen reads
\begin{equation}\label{flt}
dE=TdS-pdA+\mu dN_f,
\end{equation}
where the symbol $dA$ is not same as in the previous sections, but denotes the change of area of the
holographic screen. The above law can be understood in two different but related ways. The first is when
the holographic screen evolves with time in a time-dependent background. In such a background, for a
given time there is a local Killing vector so Verlinde temperature is well-defined, the above equation
is then about evolution of energy, entropy, area and the number of screen particles. The second interpretation
is the following. We may imagine that we move the holographic screen in space-time in a way we like,
this can be a quasi-static process (we move it very slowly), then the first law is a statement about
change of physical quantities on different screens.

The first interpretation of Eq.(\ref{flt}) is useful for a situation in cosmology, we will not discuss
this in the present paper. We will employ to second interpretation to discuss the holographic entropy
of a gas, as well as black holes.

For a static mass distribution with rotational symmetry, we consider holographic screens be spheres with
constant $r$, thus, verlinde temperature is constant everywhere on a given screen. Due to rotational symmetry,
$\sigma$, $p$ and $\mu$ are also constant on a given screen. The reason for us to consider these special
cases is that in our proposal so far, there is suggestion about the form of the chemical potential
$\mu$, but it is nevertheless non-vanishing in general.

We start with the general static, spherically symmetric metric
\begin{equation}\label{metric}
ds^{2}=-N^{2}dt^{2}+h^{2}dr^{2}+r^{2}(d\theta^{2}+\sin^{2}\theta
d\phi^{2}),
\end{equation}
where $N$ and $h$ are functions of $r$, and $\Sigma$ is a
hypersurface described by $r=$constant.

In \cite{Brown}, Brown and York suggest a subtraction procedure for the stress tensor on $\Sigma$,
since without subtraction we will get divergent results. The subtraction does not change our
derivation of the Einstein equations, since the subtraction is against a fixed background.
For an asymptotically flat
space, Brown and York propose to deduct the contributions of energy
from the flat background spacet-ime. This yields the same result at
infinity as the ADM energy  for an asymptotically flat space. We will use this procedure
for a black hole. We will also use this procedure inside a gas, since $N$ and $h$ approaches
$1$ at the origin of the gas.

Thus, we focus on spherically symmetric asymptotically flat space-time in this
paper. For more details of the subtraction scheme, please refer to
Sec. VI of \cite{Brown}. In the following, we will use the results of
\cite{Brown} without explaining their derivation.

Following the subtraction scheme used in \cite{Brown}, the surface
energy density on screen becomes
\begin{equation}
\sigma =u_{i}\xi_{j}\tau^{ij}=\frac{N}{4\pi
G}(\frac{1}{r}-\frac{1}{rh}),
\end{equation}
while the surface momentum density $j_{a}$ vanishes (due to
spherical symmetry). And the the pressure $p$ is
\begin{equation}\label{p}
p=\frac{N}{8\pi
G}(\frac{N^{\prime}}{Nh}+\frac{1}{rh}-\frac{1}{r}).
\end{equation}
where $\sigma$ and $p$ are energy density and pressure measured by
the observer at infinity which are related with those defined in
\cite{Brown} by a redshift. One can easily find that $\sigma$, $p$
vanish in flat space-time. Thus, the quasi-local energy is
\begin{equation}\label{Energy}
E=\int_{B}d^{2}x\sqrt{\sigma}\sigma =\frac{N}{G}(r-\frac{r}{h}).
\end{equation}
The Verlinde temperature on the screen is
\begin{equation}
T=\frac{\hbar}{2\pi}e^{\phi}N^{b}\nabla_{b}\phi=\frac{\hbar}{2\pi}\frac{N^{\prime}}{h}.
\end{equation}
We are ready to discuss the first law of thermodynamics on screen
\begin{equation}\label{Law1}
dE=T dS-p dA +\mu dN_{f},
\end{equation}
where $S$ is entropy on the sphere etc, the area is $A=4\pi r^2$, the above law
is a statement for adiabatically changing the radius $r$ of the sphere.

To compute the change of energy, we use Eq.(\ref{Energy}), we have
\begin{eqnarray}\label{Law2}
dE&=&\frac{1}{G}[N^{\prime}(r-\frac{r}{h})+N(1-\frac{1}{h})+N r
\frac{h^{\prime}}{h^{2}}]dr\nonumber\\
&=&-p dA+\frac{1}{8\pi G}(\frac{N h'}{h^{2}}+N')dA,
\end{eqnarray}
where we used Eq.(\ref{p}) for the expression of $p$. The second term contains
both $TdS$ and $\mu dN_f$.

To solve these two terms, let us assume that when the sphere sweeps
vacuum, $dS=0$, namely there is no heat flowing into the sphere.
Physically, this is natural since one does not expect any change of
number of states on the screen, for example a Schwarzschild black
hole has a fixed entropy. Note that in the vacuum, we have the
relation $Nh=1$, this is valid certainly for a Schwarzschild black
hole. We will assume that $N_f\sim \frac{A}{G\hbar}$, or without
loss of generality, take $N_f=\frac{A}{G\hbar}$, and sweep all
coefficients into the definition of $\mu$.

Take a Schwarzschild black hole as an example, with the metric
$ds^{2}=-(1-\frac{2GM}{r})dt^{2}+(1-\frac{2GM}{r})^{-1}dr^{2}+r^{2}(d\theta^{2}+\sin^{2}\theta
d\phi^{2})$, we have
\begin{equation}
\frac{p}{\rho}=\frac{\frac{N^{\prime}r}{(1-N)}-1}{2}=\frac{(1-\frac{GM}{r})-\sqrt{1-\frac{2GM}{r}}}{2(\sqrt{1-\frac{2GM}{r}}-(1-\frac{2GM}{r}))}.
\end{equation}
we read $\mu$ directly from Eq.(\ref{Law1}) Eq.(\ref{Law2})
\begin{eqnarray}
\mu&=&\frac{\hbar N^{\prime}}{8\pi h}(h+\frac{Nh^{\prime}}{N^{\prime}h})\nonumber\\
&=&\frac{T}{4}(h+\frac{Nh^{\prime}}{N^{\prime}h})\nonumber\\
&=&\frac{T}{4}[(h-1)+(1+\frac{Nh^{\prime}}{N^{\prime}h})-C(Nh-1)]\nonumber\\
&=&\frac{\hbar}{8\pi}(\frac{xN^{\prime}}{h}-N^{\prime}-\frac{(1-x)Nh^{\prime}}{h^{2}}-\frac{CN^{\prime}(Nh-1)}{h}),\nonumber\\
\end{eqnarray}
where in the third line we added a term $C(Nh-1)$, this term vanishes in the region of vacuum, we also
used the fact that for a vaccum region $(1+\frac{N
h^{\prime}}{N^{\prime}h}=0)$, so we obtain a general expression
\begin{equation}\label{ansm}
\mu=\frac{T}{4}(h+\frac{N h^{\prime}}{N^{\prime}h})=
\frac{T}{4}[x(h-1)+(1-x)(h+\frac{N
h^{\prime}}{N^{\prime}h})-C(Nh-1)],
\end{equation}
the above expression is certainly valid for a vacuum region, and in particular for region outside
a Schwarzschild black hole. Now, let us take a step further, assuming that the above is also valid
for a region with any kind of matter, with parameters  $x$ and $C$ to be determined. This is certainly
a stretch of logic. We think this assumption is rather reasonable for the following reason. We have learned
that according to Brown and York, both  surface energy density $\sigma$ and surface pressure $p$ are
local functions of the local geometry (here only $N$ and $h$ and their derivatives are involved), it is
natural to assume that $\mu$ is also a function of local geometry.

In fact, beside the term $C(Nh-1)$, other terms such as $C_{n}((Nh)^{n}-1)$) are
also possible. However, in our following discussion about a gas with weak gravity, this term
amounts  to redefining $C$ as  $n\ C_{n}$. From Eqs.(\ref{Law1}, \ref{Law2}, \ref{ansm}), one get the entropy
\begin{equation}\label{mr}
dS=\frac{1}{4G\hbar}(x(1+\frac{Nh'}{N'h})+C(Nh-1))dA.
\end{equation}
In the following, we are mainly interested in the holographic thermodynamics of a gas with weak gravity.
We can use this case to determine paramter $x$. This metric of this case is solved in the appendix , and we have
\begin{equation}
 ds^{2}=-(1+a r^{2}+b
r^{4})dt^{2}+(1+c r^{2}+d
r^{4})dr^{2}+r^{2}(d\theta^{2}+\sin^{2}\theta d\varphi^{2}),
\end{equation}
where $a, b, c ,d$ can be found in the appendix, and we solved the metric up to the second order in $G$, this is
needed for determining the holographic entropy of the gas. To this order, we find that two parameters of the
gas need to be introduced, namely $w_1=P(0)/\rho(0)$ and $w_2=dP/d\rho (0)$. in order not to violate the dominant
energy condition $w_1\le 1$, and there is no constraint on $w_2$.

Substituting the solution into Eq.(\ref{mr}), we find that in
general there is an area term $\sim x r^2/G\hbar$ (Eq.(\ref{S}) in
the Appendix), this is certainly too large for a gas. To make this
term absent, we take $x=0$, thus
\begin{eqnarray}
\mu&=&\frac{T}{4}[(h+\frac{N h^{\prime}}{N^{\prime}h})+C(Nh-1)] \nonumber\\
&=&\frac{\hbar}{8\pi}(N^{\prime}+\frac{Nh^{\prime}}{h^{2}}+C(NN^{\prime}-\frac{N^{\prime}}{h})),\nonumber\\
\end{eqnarray}
and
\begin{equation}\label{mmr}
 S=\frac{C \pi^{2}\rho(1+w_{1})r^{4}}{\hbar}.
\end{equation}
This is our main result in this section. This holographic entropy is
much larger than the usual entropy of a hot gas, even for the case
when $w_1=w_2=1/3$, namely radiation. The entropy increases with
$w_1$ and reaches maximum for $w_1=1$, the largest value allowed by
the dominant energy condition. This holographic entropy is quite
similar to the Bekenstein bound $2\pi Mr$. To our approximation,
$M=(4\pi/3)\rho r^3$, thus the Bekenstein bound is $(8\pi^2/3)\rho
r^4/\hbar$. Now, if we assume that the Bekenstein bound is saturated by
our holographic entropy for radiation $w_1=1/3$, then we need to
take $C=2$. If the Bekentstein bound is saturated for $w_1=1$, then
$C=4/3$. We leave determination of $C$ to a future work.

Interestingly, our result is consistent with Verlinde's entropic force. He assumes
\begin{equation}
Fdr=TdS,
\end{equation}
where $Fdr$ is the work done by gravitational force on the gas when the gas is squeezed into the holographic
screen, or the change of gravitational energy when we move the holographic screen adiabatically outward.
 Under weak gravity approximation, we have $F=C^{\prime}\rho\phi4\pi r^{2}$ so
\begin{equation}
C^{\prime}\rho\phi4\pi r^{2}dr=\frac{\hbar}{2\pi}a r dS,
\end{equation}
this leads to  $S=\frac{C^{\prime}\pi^{2}\rho r^{4}}{\hbar}$.

In retrospect, our result is quite natural. We study the situation with weak gravity expansion, the
expansion parameter is $G\rho r^2$, if the first term in the expansion of $S$ is $O(r^2/G\hbar)$,
the second must be $O(\rho r^4/\hbar)$. The interesting aspect of our discussion is that we can make
the area term dsiappear, thus the leading term is $O(\rho r^4/\hbar)$, and we predict a $w_1$ dependent
factor $1+w_1$, increasing when $w_1$ increases. It is also interesting that the form is very close
to the Bekenstein bound, thanks to the factor $\pi^2$. The undetermined numerical factor $C$ should be
a rational number, as we believe that $\mu$ is to be determined by a geometrical method. As we shall see,
the trouble with Verlinde's definition of energy is that the area term is always there.

Finally, we make a note that the holographic entropy, though much larger than the
usual statistical entropy of a gas, may help to resolve some longstanding puzzles. It has been known for a long time
that the usual statistical entropy of a collapsing system violates the second law of
thermodynamics, our explanation is that we need to use the holographic entropy rather
than the usual statistical entropy. We expect that the holographic entropy smoothly
crosses-over to the black hole entropy in a black hole formation process.

\section{A trouble with Verlinde energy}

In this section, we work with Verlinde's definition of energy to see whether we can have
a reasonable thermodynamics of a gas, we shall see that there will be always an area term, this
indicates that Verlinde energy is not a good choice.

 Verlinde's definition of energy $M$ is given by the equipartition theorem, which in turn upon using
 the Einstein equations is equal to the Komar mass or ADM mass
\begin{equation}
M=\frac{N_{f}T}{2}=\frac{AT}{2G\hbar},
\end{equation}
where $A$ is the screen area, and $N_{f}$ is the number of used bits
on screen which is supposed to be equal to the area \cite{Verlinde}, this is the same as our
choice of number of ``particles" in the previous section. The general form of
the first law of thermodynamics on the screen is
\begin{equation}
dM=T dS-p dA +\mu dN_{f}.
\end{equation}
For Schwarzschild black hole $dM=0$, $dS=0$, so we can easily derive
\begin{equation}
\mu=G\hbar
p=\frac{N\hbar}{8\pi}(\frac{h^{\prime}}{Nh}+\frac{1}{rh}-\frac{1}{r}).
\end{equation}
We choose as in the previous section
\begin{equation}\label{onec}
\mu=G\hbar p-CT(Nh-1),
\end{equation}
where the second term vanishes for a region of vacuum. One may use another function with the energy dimension
instead of $T$, we will discussion this more general choice shortly.
For a gas with weak gravity discussed in the previous section and the appendix, using the above chemical potential, one gets
\begin{equation}
dS=\frac{dM}{T}+\frac{(p-\frac{\mu}{G\hbar})dA}{T}=\frac{(CT(Nh-1)+\frac{T}{2})dA}{G\hbar
T}+\frac{A}{2G\hbar}d\ln T,
\end{equation}
where $T=\frac{2G\hbar\rho r(1+3w_{1})}{3}$ in leading order, thus
$\frac{Ad\ln T}{2G\hbar}\approx\frac{A d\ln
r}{2G\hbar}=\frac{2\pi r^{2}}{G\hbar}\frac{dr}{r}=d\frac{A}{4G\hbar}$.\\
The first term in the second equality of the above formula is of higher order in $r$ than the second
term, whihle the second term gives  $S=\frac{3A}{4G\hbar}$ in the leading order, we certainly do not expect
an area law independent of the gas energy density, this term is also way too larger than the Bekenstein bound.
We conclude that there is no reasonable holographic thermodynamics with Verlinde's choice of energy, with the choice
of the chemical potential in (\ref{onec}).

To disapprove Verlinde's energy, let us consider a more general choice of chemical potential
\begin{equation}\label{secc}
\mu=G\hbar p+ Tf(N, h, N^{(n)}, h^{(n)}),
\end{equation}
where $f$ is a function of $N$ and $h$ and their derivatives. Thus, we derive from the first law that
\begin{equation}\label{thc}
dS={dM\over T}+f(N, h, N^{(n)}, h^{(n)}){dA\over G\hbar},
\end{equation}
the first term will always give rise to an area contribution to $S$, we want to know the behavior of
$f$ for a small $r$, in this case $N, h\rightarrow 1$, $N^{(1)}, h^{(1)}\rightarrow 0$, this is the same
behavior as the vacuum, For a region of large $r$ outside of a black hole, $dS=f(1,1,0,0)dA/G\hbar$, if $f$ does not contain higher derivatives
of $N, h$ than the first order, so $f(1,1,0,0)=0$, and this implies as for a gas when $r \rightarrow 0$,
$f\rightarrow 0$ thus it is at least proportional to $r$, so the second term in (\ref{thc}) does not contain
an area term, and we can not use it to cancel the area term coming from the frist term in (\ref{thc}).
If $f$ contains, for instance, the second derivatives of $N$ and $h$, these quantities depend on $\rho$, however,
we can not use an $\rho$ dependent area term to cancel an $\rho$ independent area term.
This argument is equally applicable if even higher order derivatives are present in $f$.

We conclude that Verlinde's choice of energy is not suitable for a reasonable holographic thermodynamics.

\section{conclusion}

In the present work, we suggest a new program, in which a holographic screen is either open or closed.
The surface stress tensor is proportional to the extrinsic curvature, thus the energy we obtain is
actually Brown-York semi-local energy. This energy includes gravitational contribution.

By using the new energy definition, we are able to avoid assuming the equipartition theorem of energy.
With an open screen we do not have to face the negative temperature problem, thus the derivation
of the Einstein equations is more complete. In view of this, our proposal is more appropriate
for studying a time-dependent background, in particular cosmology.

We also discuss holographic thermodynamics of a spherically symmetric and static system, in particular
a gas of weak gravity, and obtain an interesting result for the holographic entropy. This
entropy is similar to the Bekenstein bound. This suggests that there is a tremendous contribution
to entropy from gravity. This may resolve the entropy jump puzzle in a black hole formation
process, and also may resolve violation of the second law of thermodynamics in a collapsing
gravitational system.

There remain many interesting problems. The most important one is to derive a general and geometric
formula for the chemical potential, as it is as important as the surface energy density and surface
pressure, and contains useful information for holography.

\section*{Acknowledgments}

We would like to thank Qing-Guo Huang and Yi Pang for useful discussions. This research was supported by a NSFC grant No.10535060/A050207,
a NSFC grant No.10975172, a NSFC group grant No.10821504 and Ministry of
Science and Technology 973 program under grant No.2007CB815401.

\section*{Appendix}

In this appendix we discuss an approximate solution for a gas with weak gravity. We assume the gas is spherically symmetric
and static, so the metric is
\begin{equation}
ds^{2}=-(1+a r^{2}+b r^{4})dt^{2}+(1+c r^{2}+d
r^{4})dr^{2}+r^{2}(d\theta^{2}+\sin^{2}\theta d\phi^{2}),
\end{equation}
where $a$ $b$ $c$ $d$ are coefficients to be determined. The
expansion in $r^2$ is controled by $G\rho$. By weak gravity we mean
$G\rho r^2\ll 1$. Since both the energy density and pressure are
smooth at the origin $r=0$, so there are no linear terms of $r$ in
the metric.

For expansion up to the order $O(r^4)$, the most general gas is characterized by two parameters, we
will use the following two parameters $w_{1}$ and $w_{2}$
\begin{equation}
w_{1}=\frac{P(0)}{\rho(0)}, \ \ w_{2}=\frac{dP}{d\rho}(0),
\end{equation}
where $\rho$ and $P$ are the energy density and pressure in
spacetime $M$, respectively. From the Einstein equations
\begin{equation}
G^{0}_{0}=-3c+(5c^{2}-5d)r^{2}=-8\pi G\rho(r),
\end{equation}
\begin{equation}
G^{r}_{r}=(2a-c)+(-2a^{2}+4b-2ac+c^{2}-d)r^{2}=8\pi GP(r),
\end{equation}
\begin{equation}
G^{\theta}_{\theta}=(2a-c)+(-3a^{2}+8b-3ac+2c^{2}-2d)r^{2}=8\pi
GP(r),
\end{equation}
we obtain
\begin{equation}
a = \frac{4G \pi (1 + 3 w_{1}) \rho}{3},\ \ \ \ c = \frac{8 G\pi
\rho}{3},
\end{equation}
\begin{equation}
b = \frac{(5 - \frac{1}{w_{2}})}{20 }(a^{2} + a
c)=\frac{4G^{2}\pi^{2}\rho^{2}(1+w_{1})(1+3w_{1})(5w_{2}-1)}{15w_{2}},
 \end{equation}
 \begin{equation}
 d = c^{2} -
\frac{1}{5 w_{2}} (a^{2} + a
c)=-\frac{16G^{2}\pi^{2}\rho^{2}(3(1+w_{1})(1+3w_{1})-20w_{2})}{45w_{2}},
\end{equation}
where $\rho=\rho(0)$. Note that when $w_{2}\rightarrow 0$, $b$ and $d$ will reach infinity
which implies that there is a lower limit for $w_{2}$. We also note by the way that the post-Newtonian
formalism is not appropriate for the solutions we are interested (that requires both $w$ parameters to
be very small.)

We then have
\begin{eqnarray}\label{N}
N&=&\sqrt{1 +  a r^{2}+br^{4}}\nonumber\\
&\approx&1+\frac{2G\pi\rho(1+3w_{1})r^{2}}{3}-\frac{2G^{2}\pi^{2}\rho^{2}r^{4}(3(1+w_{1})(1+3w_{1})-10w_{2}-30w_{1}w_{2})}{45w_{2}},
\end{eqnarray}
and
\begin{equation}\label{h}
h=\sqrt{1 + c r^{2} + dr^{4}}\approx1+\frac{4G\pi\rho
r^{2}}{3}+\frac{8G^{2}\pi^{2}\rho^{2}r^{4}(5w_{2}-3(1+w_{1})(1+3w_{1}))}{15w_{2}}.\
\
\end{equation}

We obtain Verlinde temperature $T$, surface pressure $p$, quqsi-local energy $E$ on the
screen $\Sigma$ and pressure $P$ in the bulk $M$ as below
\begin{eqnarray}{\label{T}}
T&=&\frac{\hbar N^{\prime}}{2\pi h}\nonumber\\
 &\approx&\frac{2G\hbar\rho(1+3w_{1})r}{3}-\frac{4G^{2}\hbar\pi\rho^{2}r^{3}(1+3w_{1})(1+w_{1})}{15w_{2}},\nonumber\\
\end{eqnarray}

\begin{eqnarray}
      p=\frac{N}{8\pi G}(\frac{N^{\prime}}{Nh}+\frac{1}{rh}-\frac{1}{r})\approx\frac{\rho w_{1} r}{2} - \frac{G \pi  r^{3}\rho^{2}(2  + 3
w_{1})}{9},
\end{eqnarray}

\begin{eqnarray}
E&=&\frac{N}{G}(r-\frac{r}{h})\nonumber\\
 &\approx&\frac{4\pi\rho r^{3}}{3}-\frac{8\pi^{2}r^{5}G\rho^{2}(3+12w_{1}+9w_{1}^{2}-10w_{2}-15w_{1}w_{2})}{45w_{2}},\nonumber\\
\end{eqnarray}
\begin{eqnarray}{\label{Pb}}
P=\rho w_{1}-\frac{2\pi G\rho^{2}}{3}(1+3w_{1})(1+w_{1})r^{2} .\ \ \ \
\ \ \ \ \ \ \ \ \ \ \  \ \ \ \ \ \ \ \ \ \ \ \
\end{eqnarray}
$\rho (r)$ can be obtained from $P$ by the definition of two $w$ parameters.

The positivity of  temperature (\ref{T}) requires $w_1>-1/3$, namely the gas generate attractive force
only, requiring and bulk pressure (\ref{Pb}) be
positive, we have $w_1\ge 0$,  but this is not a physics condition. The Dominant Energy Condition
($\rho\geq \ \mid P\mid$) implies $w_1\le 1$, so the physical range of $w_1$ is $-1/3 < w_1\le 1$.

Now, we investigate the first law of thermodynamics on the screen to
derive the holographic entropy of the gas:
\begin{equation}
dE=T dS-p dA +\mu dN_{f}.
\end{equation}
As discussed in Sec. IV, the general expression of $\mu$ is
\begin{equation}
\mu=\frac{T}{4}(h+\frac{N h^{\prime}}{N^{\prime}h})=
\frac{T}{4}[x(h-1)+(1-x)(h+\frac{N
h^{\prime}}{N^{\prime}h})-C(Nh-1)],
\end{equation}
from which we have
\begin{equation}
dS=\frac{1}{4G\hbar}(x(1+\frac{Nh'}{N'h})+C(Nh-1))dA.
\end{equation}
Using Eq.(\ref{N}) and Eq.(\ref{h}), we obtain
\begin{eqnarray}\label{S}
S&=&x(\frac{3\pi
r^2}{G\hbar}\frac{1+w_{1}}{1+3w_{1}}+\frac{2\pi^2\rho
r^4}{15\hbar}\frac{(3w_1-1)(3(1+w_1)(1+3w_1)-5w_2(5+3w_1))}{(1+3w_1)^2
w_{2}})\nonumber\\
&+&\frac{C \pi^{2}\rho(1+w_{1})r^{4}}{\hbar}.
\end{eqnarray}
The first term $x\frac{3\pi r^2}{G\hbar}\frac{1+w_{1}}{1+3w_{1}}$ is
too large for gas, so to make this term absent, we take $x=0$, thus
\begin{eqnarray}
S=\frac{C \pi^{2}\rho(1+w_{1})r^{4}}{\hbar}.
\end{eqnarray}

\end{document}